\documentclass[12pt]{iopart}

\usepackage{latexsym}
\usepackage{amssymb}
\usepackage{multirow}
\usepackage{amsthm}
\usepackage{amsfonts}
\usepackage{graphicx,color}
\usepackage{bm}
\usepackage[hypertex]{hyperref}
\makeatletter

\newcommand{\Rmnum}[1]{\expandafter\@slowromancap\romannumeral #1@}
\makeatother

\begin{document}

\title[]
{Symmetry analysis of transport properties in helical superconductor junctions}

\author{Qiang Cheng$^1$, Yinhan Zhang$^2$, Kunhua Zhang$^3$, Biao Jin$^4$ and Changlian Zhang$^1$}

\address{$^1$ School of Science, Qingdao Technological University, Qingdao, Shandong 266520, China}
\address{$^2$ International Center for Quantum Materials, Peking University, Beijing 100871, China}
\address{$^3$ ICQD, Hefei National Laboratory for Physical Sciences at Microscale, University of Science and Technology of China,
Hefei, Anhui 230026, China}
\address{$^4$ School of Physics, University of Chinese Academy of Sciences, Beijing 100049, China}

\ead{\mailto{chengqiang07@mails.ucas.ac.cn}}
\ead{zhangkh1@ustc.edu.cn}
\ead{biaojin@ucas.ac.cn}
\date{\today}

\begin{abstract}
We study discrete symmetries satisfied by helical $p$-wave superconductors with d-vectors $k_{x}\hat{x}\pm k_{y}\hat{y}$
or $k_{y}\hat{x}\pm k_{x}\hat{y}$ and transformations brought by the symmetry operations to ferromagnet and spin-singlet superconductors, which show intimate associations with transport properties in heterojunctions including helical superconductor. Especially, the partial symmetries of the Hamiltonian under the spin-rotation and gauge-rotation operations are responsible for novel invariances of the conductance in tunnel junctions and new selection rules of the lowest current and peculiar phase diagrams in
Josephson junctions which are reported recently. The symmetries of constructed free energies for Josephson junctions are also analyzed which are consistent with the results from Hamiltonian.
\end{abstract}

\maketitle

\tableofcontents
\clearpage

\section{Introduction}
Symmetries play an important role in the classification of topological matters \cite{Hasan,Qi,Tanaka,Ando,Chiu,Sato} and the definition of topological invariants \cite{Kane,Sato2,Sato3,Kimme,Wakatsuki,Ezawa}. The spin-triplet superconductors (TSs) with d-vectors $k_{x}\hat{x}\pm k_{y}\hat{y}$ or $k_{y}\hat{x}\pm k_{x}\hat{y}$ belong to symmetry class DIII according to the tenfold classification of topological insulators and superconductors \cite{Ryu,Schnyder,Schnyder2}, which satisfy time-reversal symmetry (TRS), particle-hole (or charge-conjugation) symmetry (PHS) and chiral symmetry (CS) but break SU(2) spin-rotation symmetry (SRS). The superconducting systems, possessing the non-trivial $\mathbb{Z}_{2}$ topological invariant and supporting helical Majorana modes, are analogous to the quantum spin Hall state \cite{Kane2,Bernevig}. Such superconductors are called helical superconductors. The helical superconducting states are candidates for paring in Sr$_2$RuO$_{4}$ \cite{Mackenzie,Maeno,Zhang} and the triplet part of the oder parameter in the non-centrosymmetric superconductor CePt$_{3}$Si \cite{Bauer}. Among others, $k_{x}\hat{x}+k_{y}\hat{y}$ is the two-dimensional analog of the Balian-Werthamer state (B phase) in $^3$He \cite{Mizushima}. In contrast, the chiral TSs with d-vectors $(k_{x}\pm k_{y})\hat{z}$ can be viewed as two copies of the spinless superconductor which belongs to the symmetry D class with the non-trivial $\mathbb{Z}$ topological number and a chiral Majorana mode \cite{Qi,Tanaka,Chiu}. The chiral superconducting states satisfy PHS and SRS about the $\hat{z}$-axis but beak TRS, which are analogous to the quantum Hall state \cite{Furusaki}.

Researches on transport properties of topological junctions not only can provide methods for the detection of
the Majorana fermion edge states \cite{Fu,Zuo,Tanaka2,Lutchyn,Linder,Asano,Beenakker1,Beenakker2} but also can probe the symmetries satisfied by the topological superconductors \cite{Kashiwaya,Lu,Tanaka3}. The charge conductance in the ferromagnet (F)$\vert$chiral TS junction is invariant when the magnetization in F is rotated about the $\hat{z}$-aixs due to the SU(2) SRS kept by TS \cite{Hirai}. The Josephson effects in spin-singlet superconductor (SS)$\vert$F$\vert$chiral TS junctions show that the form of the lowest current ($\sin{\phi}$ or $\cos{\phi}$) are strongly related to the symmetries of Cooper pair functions in SS \cite{Brydon1}. The $0$-$\pi$ phase transition in the purely chiral superconductor junctions with different chiralities are also clarified \cite{Brydon2}. These results are very different from those for the helical superconductor junctions since the SU(2) rotation symmetry for the latter is completely broken. The conductance in F$\vert$helical TS junctions exhibits higher anisotropy when the direction of magnetization is changed \cite{Cheng1}. Both $\sin{\phi}$-type and $\cos{\phi}$-type of current in the SS$\vert$F$\vert$helical TS junctions always exists; the selection rules for the current can be given by the orientation of magnetization \cite{Cheng2}. The purely helical Josephson junctions with different d-vectors can host more ground states such as $\pi$-phase, $\varphi$-phase and $\varphi_{0}$-phase which can be tuned easily by the direction of magnetization \cite{Cheng3}. Especially, the $\varphi_{0}$-phase is recently receiving an increasing interest \cite{Zuo,Krive,Dolcini,Reynoso,Zazunov,Yokoyama,Dolcini2,Szombati}.

However, the SU(2) symmetry breaking in helical superconductor junctions does not mean the absence of symmetries of transport quantities. Actually, both the charge conductance and the Josephson current possess invariances under the rotation of magnetization \cite{Cheng1}. The invariances have important effects to classify the SS$\vert$F$\vert$helical TS junctions into two types and to form the symmetric phase diagrams of the purely helical junctions \cite{Cheng2,Cheng3}. In this paper, we consider the transport properties of the F$\vert$helical TS, SS$\vert$F$\vert$helical TS and helical TS$\vert$F$\vert$helical TS junctions from the viewpoint of symmetries of Hamiltonian in order to clarify the origin of the invariances of transport quantities. The main findings in this paper are as following.

(a) The Hamiltonian for helical superconductor is invariant under the joint operation of the SU(2) spin-rotation and the U(1) gauge transformation, which will lead to the invariances of conductance and Josephson current.

(b) The Hamiltonian satisfies partial symmetries under particular spin-rotation operation or the joint operation of rotation and gauge transformation; the operations can transform the BdG Hamiltonian from $H(k_{x},k_{y})$ into $H(k_{x},-k_{y})$. The partial symmetries are responsible for the symmetries of transport quantities and phase diagrams. Especially, if we take the different $k_{y}$ parities of superconducting wave functions in SSs into account, the partial symmetries will produce different selection rules of the lowest order current for different SS$\vert$F$\vert$helical TS junctions, which will divide the junctions into two types.

(c) TRS and PHS of the Hamiltonian bring about the sign reversal of Josephson current and the symmetry of the current for junctions without the non-magnetic interfacial potential, respectively.

(d) The symmetries of current derived from the Hamiltonian are consistent with those from the constructed free energies for the helical Josephson junctions, which directly reflects the mechanism of interaction between ferromagnetism and helical superconductivity.

\section{BdG Hamiltonian}

\subsection{Superconducting system}
The generic Hamiltonian for superconductor is
\begin{eqnarray}\label{Ham}
\mathcal{H}=\frac{1}{2}\sum_{{\bf{k}}}\Psi_{{\bf{k}}}^{+}H({\bf{k}})\Psi_{{\bf{k}}},
\end{eqnarray}
with $\Psi_{{\bf{k}}}=(c_{{\bf{k}}\uparrow},c_{{\bf{k}}\downarrow},c_{{-\bf{k}}\uparrow}^{+},c_{{-\bf{k}}\downarrow}^{+})^{T}$ and the BdG Hamiltonian
\begin{eqnarray}\label{bdg}
H({\bf{k}})=\left(
\begin{array}{cc}
\hat{\epsilon}({\bf{k}})&\hat{\Delta}({\bf{k}})\\
-\hat{\Delta}^{*}({-\bf{k}})&-\hat{\epsilon}^{*}(-{\bf{k}})
\end{array}\right)£¬
\end{eqnarray}
where $c_{{\bf{k}}\alpha}$($c^{+}_{{\bf{k}}\alpha}$) is the annihilation (creation) operator of electron with momentum ${\bf{k}}=(k_{x},k_{y})$ and spin $\alpha$, $\hat{\epsilon}({\bf{k}})$ describes the normal dispersion of electron and $\hat{\Delta}({\bf{k}})$ denotes the energy gap matrix. The normal state is invariant under the space inversion, i.e., $\hat{\epsilon}({\bf{k}})=\hat{\epsilon}({-\bf{k}})$.
\subsubsection{Spin-triplet system}
\begin{figure}[h!]
\begin{center}
\includegraphics[width=5cm]{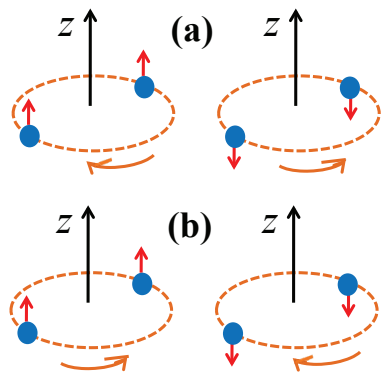}
\end{center}
\caption{(Color online) Schematic illustration of Cooper pairs for the helical states which are superposition of equal-spin wave functions with $S_{z}=\pm1$. (a) For the $p_{xy}^{+}$ and $p_{yx}^{-}$-wave, the orbital angular momentum $(l,l_{z})$ for the spin-up part with $S_{z}=1$ and the spin-down part with $S_{z}=-1$ is $(1,-1)$ and $(1,1)$, respectively. (b) For the $p_{xy}^{-}$ and $p_{yx}^{+}$-wave, the angular momentum $(l,l_{z})$ for the spin-up part with $S_{z}=1$ and the spin-down part with $S_{z}=-1$ is $(1,1)$ and $(1,-1)$, respectively. The $z$-axis has been chosen along the crystallographic $c$-axis.}
\label{f1}
\end{figure}
For TS, the matrix $\hat{\Delta}({\bf{k}})=({\bf{d}}({\bf{k}})\cdot\hat{\sigma})i\hat{\sigma}_{2}$ with $\hat{\Delta}({-\bf{k}})=-\hat{\Delta}({\bf{k}})$ in which $\hat{\sigma}_{i}$ ($i=1,2,3$) denote the usual Pauli matrices in spin space and ${\bf{d}}({\bf{k}})$ the d-vector. The helical TS with the $d$-vector ${\bf{d}}_{uv}^{\pm}({\bf{k}})=\Delta_{0}(k_{u}\hat{x}\pm k_{v}\hat{y})$ with $(u,v=x,y)$ is denoted by $p_{uv}^{\pm}$-wave TS respectively in this paper which can be regarded as the superposition of the equal-spin wave functions with different orbital angular momentum,
\begin{eqnarray}
\mbox{p}_{xy}^{+}:~~-(k_{x}-ik_{y})\mid\uparrow\uparrow\rangle+(k_{x}+ik_{y})\mid\downarrow\downarrow\rangle,\\
\mbox{p}_{xy}^{-}:~~-(k_{x}+ik_{y})\mid\uparrow\uparrow\rangle+(k_{x}-ik_{y})\mid\downarrow\downarrow\rangle,\\
\mbox{p}_{yx}^{+}:~~i(k_{x}+ik_{y})\mid\uparrow\uparrow\rangle+i(k_{x}-ik_{y})\mid\downarrow\downarrow\rangle,\\
\mbox{p}_{yx}^{-}:~~-i(k_{x}-ik_{y})\mid\uparrow\uparrow\rangle-i(k_{x}+ik_{y})\mid\downarrow\downarrow\rangle,
\end{eqnarray}
as shown in figure 1. Note, the phase before the orbital wave functions is different for $p_{uv}^{\pm}$-wave states. Actually, the Hamiltonian for the $p_{uv}^{\pm}$-wave states can be written in a block diagonal form in the basis $\Psi=(c_{{\bf{k}}\uparrow},c_{{-\bf{k}}\uparrow}^{+},c_{{\bf{k}}\downarrow},c_{{-\bf{k}}\downarrow}^{+})^{T}$ which is
\begin{eqnarray}
H({\bf{k}})=\left(
\begin{array}{cccc}
\epsilon({\bf{k}})&-d_{x}+id_{y}&0&0\\
-d_{x}-id_{y}&-\epsilon({\bf{k}})&0&0\\
0&0&\epsilon({\bf{k}})&d_{x}+id_{y}\\
0&0&d_{x}-id_{y}&-\epsilon({\bf{k}})
\end{array}\right).
\end{eqnarray}
The block diagonal Hamiltonian in the spin subspace is similar to the one discussed in \cite{Qi2} which describes the system consisting of two separate copies of spinless superconductors with the opposite chirality. That is the reason for our use of the term `helical superconductor' in this paper.

The helical superconductor possesses TRS. We introduce the time-reversal operator $\mathcal{T}=\chi_{T} K$ where $\chi_{T}=\hat{\tau}_{0}\otimes i\hat{\sigma}_{2}$ with $\hat{\tau}_{0}$ the $2\times2$ unit matrix in particle-hole space and $K$ the complex conjugation operator, which flips the sign of momentum and the direction of spin. The actions of $\mathcal{T}$ on the creation and the annihilation operators give
$\mathcal{T}c_{{\bf{k}}\alpha}\mathcal{T}^{-1}=\sum_{\alpha'}i({\hat{\sigma}_{2}})_{\alpha\alpha'}c_{-{\bf{k}}\alpha'}$ and $\mathcal{T}c_{{\bf{k}}\alpha}^{+}\mathcal{T}^{-1}=\sum_{\alpha'}c_{-{\bf{k}}\alpha'}^{+}i({\hat{\sigma}_{2}})_{\alpha'\alpha}^{T}$, respectively. The symmetry with $[\mathcal{T},\mathcal{H}]=0$ requires
\begin{eqnarray}\label{TS}
\chi_{T}H({\bf{k}})\chi_{T}^{-1}=H^{*}({-\bf{k}}),
\end{eqnarray}
which is obviously satisfied by the BdG Hamiltonian. After the time-reversal operation, the Hamiltonian in equation (\ref{bdg}) keeps invariant. However, if the superconductor has a phase $\phi$ which is an important quantity in the Josephson effect, the operation $\mathcal{T}$ will change the phase into $-\phi$.

The superconductor also obeys PHS. We define the charge conjugation operator $\mathcal{C}=\chi_{C}K$ with $\chi_{C}=\hat{\tau}_{1}\otimes\hat{\sigma}_{0}$ where $\hat{\tau}_{i}$ is Pauli matrices in particle-hole space and $\hat{\sigma}_{0}$ the unit matrix in spin space. The action of $\mathcal{C}$ on the electron state $\mid{\bf{k}}\uparrow\rangle$ will produce a hole in the state $\mid-{\bf{k}}\downarrow\rangle$. Consequently, the transformations of the creation and the annihilation operators are $\mathcal{C}c_{{\bf{k}}\alpha}\mathcal{C}^{-1}=c_{{\bf{k}}\alpha}^{+}$ and $\mathcal{C}c_{{\bf{k}}\alpha}^{+}\mathcal{C}^{-1}=c_{{\bf{k}}\alpha}$, respectively.
The symmetry with $\{\mathcal{C},\mathcal{H}\}=0$ requires
\begin{eqnarray}\label{PHS}
\chi_{C}H({\bf{k}})\chi_{C}^{-1}=-H^{*}({-\bf{k}}),
\end{eqnarray}
which is also satisfied by the BdG Hamiltonian in equation (\ref{bdg}). After the charge conjugation operation, the energy of quasiparticles becomes negative. In other words, $\hat{\epsilon}({\bf{k}})$ and $\hat{\Delta}(\bf{k})$ in equation (\ref{bdg}) become $-\hat{\epsilon}(-\bf{k})$ and $\Delta(-\bf{k})$, respectively. However, the phase $\phi$ still keeps its value.

Since the system satisfies both TRS and PHS, we can define a CS operator $\mathcal{S}=-iTC=\hat{\tau}_{1}\otimes\hat{\sigma}_{2}$ which gives $\mathcal{S} c_{\bf{k}\alpha}\mathcal{S}^{-1}=\sum_{\alpha'}c_{{-\bf{k}}\alpha'}^{+}(i\hat{\sigma}_{2})_{\alpha'\alpha}^{T}$ and $\mathcal{S} c_{\bf{k}\alpha}^{+}\mathcal{S}^{-1}=\sum_{\alpha'}(i\hat{\sigma}_{2})_{\alpha\alpha'}c_{-{\bf{k}}\alpha'}$. A Hamiltonian will possess CS, $\{\mathcal{S},\mathcal{H}\}=0$, if
\begin{eqnarray}\label{CS}
\mathcal{S} H({\bf{k}})\mathcal{S}^{-1}=-H({\bf{k}}).
\end{eqnarray}
Evidently, the BdG Hamiltonian for the helical superconductor has this property. After the chiral symmetry operation, $\hat{\epsilon}({\bf{k}})$, $\hat{\Delta}(\bf{k})$ and $\phi$ in equation (\ref{bdg}) will become $-\hat{\epsilon}(-\bf{k})$, $\Delta(-\bf{k})$ and $-\phi$.

Finally, we consider the spin rotation operation. The rotation matrix in the particle-hole space is diagonal, i.e., $\mathcal{R}(\xi,\eta)=\mbox{diag}(\hat{R}(\xi,\eta),\hat{R}^{*}(\xi,\eta))$ with the element
\begin{eqnarray}\label{SOM}
\hat{R}(\xi,\eta)=\left(
\begin{array}{cc}
\cos{\frac{\xi}{2}}e^{-i\frac{\eta}{2}}&-\sin{\frac{\xi}{2}}e^{-i\frac{\eta}{2}}\\
\sin{\frac{\xi}{2}}e^{i\frac{\eta}{2}}&\cos{\frac{\xi}{2}}e^{i\frac{\eta}{2}}
\end{array}\right),
\end{eqnarray}
in the spin space. The element $\hat{R}$ represents a $\xi$-angle rotation about the $y$-axis followed by a $\eta$-angle rotation about the $z$-axis. For a system with SRS, $[\mathcal{R},\mathcal{H}]=0$, one has
\begin{eqnarray}\label{SOS}
\mathcal{R}(\xi,\eta)H({\bf{k}})\mathcal{R}(\xi,\eta)^{-1}=H({\bf{k}}),
\end{eqnarray}
for all values of $\xi$ and $\eta$,
which is not satisfied by the Hamiltonian with ${\bf{d}}_{uv}^{\pm}({\bf{k}})=\Delta_{0}(k_{u}\hat{x}\pm k_{v}\hat{y})$. Actually, the helical superconductor breaks all the non-trivial rotation symmetries with the angle $\xi\ne0$ and $\eta\ne0$. Nevertheless, there are three special rotations, $\mathcal{R}(0,\pi), \mathcal{R}(\pi,0)$ and $\mathcal{R}(\pi,\pi)$, which can help bring the so-called partial symmetries.

For the $p_{xy}^{\pm}$-wave TS with d-vector ${\bf{d}}_{xy}^{\pm}({\bf{k}})$, we have
\begin{eqnarray}
\mathcal{R}'(0,\pi)H({\bf{k}})\mathcal{R}'(0,\pi)^{-1}=H({\bf{k}}),\label{gr1}\\
\mathcal{R}'(\pi,0)H({\bf{k}})\mathcal{R}'(\pi,0)^{-1}=H(k_{x},-k_{y}),\label{gr2}\\
\mathcal{R}(\pi,\pi)H({\bf{k}})\mathcal{R}(\pi,\pi)^{-1}=H(k_{x},-k_{y}), \label{pr}
\end{eqnarray}
where the operation $\mathcal{R}'(\xi,\eta)\equiv \mbox{diag}(\hat{R}(\xi,\eta)U_{1}(\frac{\pi}{2}),\hat{R}^{*}(\xi,\eta)U_{1}^{*}(\frac{\pi}{2}))$ with the pure rotation $\hat{R}(\xi,\eta)$ and the $\frac{\pi}{2}$-$U_{1}$ gauge transformation, which will be called `gauge-rotation' operation for simplicity. Equation (\ref{gr1}) shows the Hamiltonian is invariant under the gauge-rotation $\mathcal{R}'(0,\pi)$; equations (\ref{gr2}) and (\ref{pr}) indicate the Hamiltonian obeys the partial symmetry under the gauge-operation $\mathcal{R}'(\pi,0)$ or the pure rotation $\mathcal{R}(\pi,\pi)$.

For the $p_{yx}^{\pm}$-wave TS with d-vector ${\bf{d}}_{yx}^{\pm}({\bf{k}})$, equation (\ref{gr1}) still holds. Equations (\ref{gr2}) and (\ref{pr}) turn into
\begin{eqnarray}
\mathcal{R}(\pi,0)H({\bf{k}})\mathcal{R}(\pi,0)^{-1}=H(k_{x},-k_{y}),\label{pr2}\\
\mathcal{R}'(\pi,\pi)H({\bf{k}})\mathcal{R}'(\pi,\pi)^{-1}=H(k_{x},-k_{y}),\label{gr}
\end{eqnarray}
respectively. Similarly, the Hamiltonian satisfies the partial symmetries. In writing equations (\ref{gr1})-(\ref{gr}), we have used the even parity of $\hat{\epsilon}({\bf{k}})$ and the odd parity of $\hat{\Delta}({\bf{k}})$. Obviously, after the rotation operations $\mathcal{R}'(\pi,0)$ and $\mathcal{R}(\pi,\pi)$ ($\mathcal{R}(\pi,0)$ and $\mathcal{R}'(\pi,\pi)$) for the $p_{xy}^{\pm}$($p_{yx}^{\pm}$)-wave TS, $\hat{\Delta}(\bf{k})$ in equation (\ref{bdg}) will become $\hat{\Delta}(k_{x},-k_{y})$.

\subsubsection{Spin-singlet system}
For SS, the gap matrix $\hat{\Delta}({\bf{k}})=\Delta({\bf{k}})i\sigma_{2}$ with the even parity $\hat{\Delta}({\bf{k}})=\hat{\Delta}({-\bf{k}})$ in equation (\ref{bdg}). The energy gap function $\Delta({\bf{k}})=\Delta_{0}$ for the $s$-wave pairing, $\Delta({\bf{k}})=\Delta_{0}(k_{x}^{2}-k_{y}^{2})$ for the $d_{x^2-y^2}$-wave paring and $\Delta({\bf{k}})=\Delta_{0}2k_{x}k_{y}$ for the $d_{xy}$-wave paring. The spin-singlet system not only preserves TRS, PHS and CS but also obeys SRS. In order to conveniently discuss the symmetries of transport quantities later, here we give the changes of $\hat{\epsilon}({\bf{k}})$ and $\hat{\Delta}(\bf{k})$ in equation (\ref{bdg}) after the operations $\mathcal{T}$, $\mathcal{C}$ and $\mathcal{S}$,
\begin{eqnarray}
\mathcal{T}:~~\hat{\epsilon}({\bf{k}}),\hat{\Delta}({\bf{k}}),\phi\rightarrow\hat{\epsilon}({\bf{k}}),\hat{\Delta}({\bf{k}}),-\phi;
\label{TSS}\\
\mathcal{C}:~~\hat{\epsilon}({\bf{k}}),\hat{\Delta}({\bf{k}}),\phi\rightarrow-\hat{\epsilon}({-\bf{k}}),\hat{\Delta}({-\bf{k}})
e^{i\pi},\phi;\label{CSS}\\
\mathcal{S}:~~\hat{\epsilon}({\bf{k}}),\hat{\Delta}({\bf{k}}),\phi\rightarrow-\hat{\epsilon}({-\bf{k}}),\hat{\Delta}({-\bf{k}})
e^{i\pi},-\phi.\label{SSS}
\end{eqnarray}

\begin{figure}[h!]
\begin{center}
\includegraphics[width=7cm]{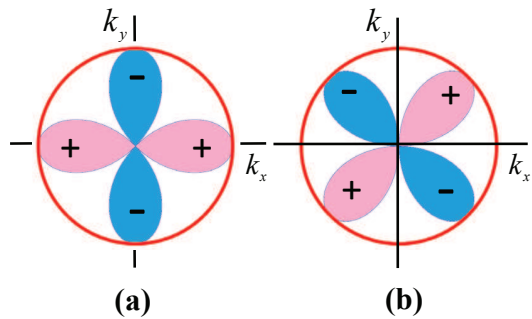}
\end{center}
\caption{(Color online) The gap functions for (a) the $d_{x^2-y^2}$-wave SS and (b) the $d_{xy}$-wave SS. The former is even about $k_{x}$ or $k_{y}$ while the latter is odd. The red circle denotes the gap function of the isotropic $s$-wave SS.}
\label{f2}
\end{figure}

\begin{table}[h!]
\renewcommand\arraystretch{1.3}
\caption{Changes of the gap matrix $\hat{\Delta}(\bf{k})$ for SS after the rotation operations.
}
\label{t1}
\begin{center}
\begin{small}
\begin{tabular}{c|c|c|c|c|c}
\hline
&$\mathcal{R}'(0,\pi)$&$\mathcal{R}'(\pi,0)$&$\mathcal{R}(\pi,\pi)$&$\mathcal{R}(\pi,0)$&$\mathcal{R}'(\pi,\pi)$\\\hline
$s, d_{x^2-y^2}$ & $e^{i\pi}\hat{\Delta}({\bf{k}})$ & $e^{i\pi}\hat{\Delta}(k_{x},-k_{y})$ & $\hat{\Delta}(k_{x},-k_{y})$ & $\hat{\Delta}(k_{x},-k_{y})$ & $e^{i\pi}\hat{\Delta}(k_{x},-k_{y})$\\ \hline
$d_{xy}$ & $e^{i\pi}\hat{\Delta}({\bf{k}})$ & $\hat{\Delta}(k_{x},-k_{y})$ & $e^{i\pi}\hat{\Delta}(k_{x},-k_{y})$ & $e^{i\pi}\hat{\Delta}(k_{x},-k_{y})$ & $\hat{\Delta}(k_{x},-k_{y})$\\ \hline
\end{tabular}
\end{small}
\end{center}
\end{table}
The changes brought by the rotation operations for SS are presented in table \ref{t1}. For later use, the transformed gap matrix has been expressed with $(k_{x},-k_{y})$ in order to keep consistent with equations (\ref{gr2})-(\ref{gr}). When we do this, there will be a $\pi$-phase difference between the gap matrix for $s$($d_{x^2-y^2}$)-wave and that for $d_{xy}$-wave. The difference originates from their opposite `parities' under $k_{y}\rightarrow-k_{y}$ as shown in figure \ref{f2}, although they are both spin-singlet even parity superconductors. The subtle difference between the gap functions can bring important physical results of Josephson effects which will seen in part 4. For simplicity, $\hat{\epsilon}(\bf{k})$ is omitted in the table since it is invariant under the operations and it is a even function both for $k_{x}$ and $k_{y}$.

\subsection{Ferromagnetic system}
The BdG-type Hamiltonian in equation (\ref{Ham}) for F is denoted by $H_{F}({\bf{k}})$ which can be written as
\begin{eqnarray}\label{HF}
H_{F}({\bf{k}})=\left(
\begin{array}{cc}
\hat{\epsilon}({\bf{k}})-{\bf{M}}\cdot\hat{\sigma}+\hat{V}_{0}&0\\
0&-\hat{\epsilon}^{*}({-\bf{k}})+{\bf{M}}\cdot\hat{\sigma}^{*}-\hat{V}_{0}
\end{array}\right),
\end{eqnarray}
where the magnetization ${\bf{M}}=M\hat{{\bf{n}}}(\theta_{m},\phi_{m})$ is specified by the direction $\hat{\bf{n}}=(\sin\theta_{m}\cos{\phi_{m}}, \sin{\theta_{m}}\sin{\phi_{m}}, \cos{\theta_{m}})$ with the polar angle $\theta_{m}$ and the azimuthal angle $\phi_{m}$; $\hat{V}_{0}=V_{0}\hat{1}_{2\times2}$ is the non-magnetic potential. The Hamiltonian $H_{F}({\bf{k}})$ can describe the bulk F or the ferromagnetic interface in heterostructures. We take $\hat{V}_{0}=0$ for the bulk and the purely ferromagnetic interface cases.

The ferromagnetic system breaks TRS, PHS and SRS, i.e., the Hamiltonian in equation (\ref{HF}) does not satisfy equations (\ref{TS})-(\ref{CS}). For later use, we also present the changes of the Hamiltonian brought by the $\mathcal{T}$, $\mathcal{C}$ and $\mathcal{S}$ transformations as what we have done for SS,
\begin{eqnarray}
\mathcal{T}:~~\hat{\epsilon}({\bf{k}}),\hat{V}_{0},\hat{{\bf{n}}}(\theta_{m},\phi_{m})\rightarrow
\hat{\epsilon}({\bf{k}}),\hat{V}_{0},\hat{{\bf{n}}}(\pi-\theta_{m},\pi+\phi_{m});\label{TF}\\
\mathcal{C}:~~\hat{\epsilon}({\bf{k}}),\hat{V}_{0},\hat{{\bf{n}}}(\theta_{m},\phi_{m})\rightarrow
-\hat{\epsilon}({-\bf{k}}), -\hat{V}_{0},\hat{{\bf{n}}}(\pi-\theta_{m},\pi+\phi_{m});\label{CF}\\
\mathcal{S}:~~\hat{\epsilon}({\bf{k}}),\hat{V}_{0},\hat{{\bf{n}}}(\theta_{m},\phi_{m})\rightarrow
-\hat{\epsilon}({-\bf{k}}),-\hat{V}_{0},\hat{{\bf{n}}}(\theta_{m},\phi_{m}).\label{SF}
\end{eqnarray}
From the above equations, it is found the direction of magnetization is `rotated' after $\mathcal{T}$ and $\mathcal{C}$ operations; the energy $\hat{\epsilon}(\bf{k})$ and $\hat{V}_{0}$ become negative after the $\mathcal{S}$ operation.

\begin{table}[h]
\begin{center}
\caption{The direction of magnetization after rotation operations.}
\label{t2}
\begin{small}
\begin{tabular}{c|c|c}
\hline
$\mathcal{R}(0,\pi),\mathcal{R}'(0,\pi)$&$\mathcal{R}(\pi,0),\mathcal{R}'(\pi,0)$&$\mathcal{R}(\pi,\pi),\mathcal{R}'(\pi,\pi)$\\\hline
$\hat{{\bf{n}}}(\theta_{m},\pi+\phi_{m})$&$\hat{{\bf{n}}}(\pi-\theta_{m},\pi-\phi_{m})$&$\hat{{\bf{n}}}(\pi-\theta_{m},2\pi-\phi_{m})$\\
\hline
\end{tabular}
\end{small}
\end{center}
\end{table}
The actions of rotation operations are summarized in table \ref{t2}. A pure rotation and its corresponding gauge-rotation lead to the same change of the ferromagnetic system due to the $U_{1}$ gauge symmetry obeyed by F. For simplicity, we do not show $\hat{\epsilon}(\bf{k})$ and $\hat{V}_{0}$ in the table since they are invariant under rotation operations.

\section{Symmetries of charge conductance}
\subsection{Blonder-Tinkham-Klapwijk formalism}
\begin{figure}[h!]
\begin{center}
\includegraphics[width=7cm]{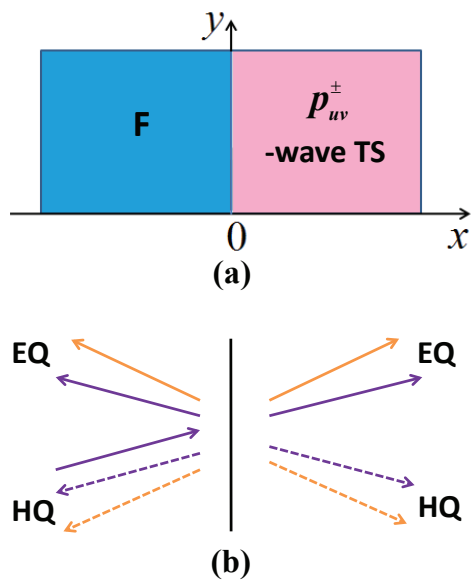}
\end{center}
\caption{(Color online) (a) Schematic illustration of the F$\vert$helical TS junctions. The current is
flowing along the $x$-axis which is defined by the crystallographic $a$-axis. (b) The scattering process for an electron-like quasiparticle (EQ) injected from F. The solid lines represent the normally reflected and the transmitted EQs. The dashed lines represent the Andreev reflected and the transmitted hole-like quasiparticles (HQs).  The spin for the injected EQ is assumed along $\hat{\bf{n}}(\theta_{m},\phi_{m})$; quasiparticles denoted by lines of the same color possess the same spin.}
\label{f3}
\end{figure}
We consider the F$\vert$helical $p$-wave TS junctions as shown in figure \ref{f3}(a). The non-magnetic interface with $M=0$ is located at $x=0$ and along $y$ axis. The ferromagnetic region with $x<0$ is described by the Hamiltonian in equation (\ref{HF}) with $V_{0}=0$; the superconducting region with $x>0$ is described by the Hamiltonian in equation (\ref{bdg}). The wave functions of quasiparticles in F and TS can be obtained through solving the BdG equations $H_{F}({\bf{k}})\Psi=E({\bf{k}})\Psi$ and $H({\bf{k}})\Psi=E({\bf{k}})\Psi$, respectively.

For the spin of quasiparticles in F, we chose the direction $\hat{\bf{n}}(\theta_{m},\phi_{m})$ as the quantization axis.
When an electron with wavevector $k_{y}$ is injected from F, there will be four physical processes as shown in figure \ref{f3}(b). The electron is normally reflected as electrons and Andreev reflected as holes; the electron transmits into TS as electron-like quasiparticles and hole-like quasiparticles. The wave function in F (TS) is the superposition of the reflected (transmitted) quasiparticles. The reflection and the transmission coefficients of the quasiparticles can be derived from boundary conditions at the interface, which are functions of $M, V_{0}, \hat{\bf{n}}(\theta_{m},\phi_{m}), k_{y}$ and the bias $V$. According to the Blonder-Tinkham-Klapwijk formalism \cite{BTK}, the angle-resolved conductance $\sigma'$ of the junctions can be expressed as the composition of the coefficients, which is a function of $\theta_{m},\phi_{m}$ and $k_{y}$ when $M, V_{0}$ and $V$ are fixed.
As a result, the angle-averaged conductance $\sigma$ can be given by
\begin{eqnarray}
\sigma(\theta_{m},\phi_{m})=C_{0}\sum_{k_{y}}\sigma'(\theta_{m},\phi_{m},k_{y}),\label{Con}
\end{eqnarray}
where $C_{0}$ is a constant independent of $\theta_{m},\phi_{m}$ and $k_{y}$. Generally speaking, the expression of the angle-averaged conductance is very complex. A more detailed description for the derivation of conductance can be found in Appendix.

\subsection{F$\vert$helical TS junctions}
Now, we clarify the symmetries of the conductance. We want to know which orientations of magnetization can lead to the same conductance. First, if the superconductor is invariant under some unitary operation and simultaneously the operation rotates the direction of magnetization from $\hat{\bf{n}}(\theta_{m},\phi_{m})$ to $\hat{\bf{n}}(\theta'_{m},\phi'_{m})$, we will have $\sigma(\theta_{m},\phi_{m})=\sigma(\theta'_{m},\phi'_{m})$. For SS, the spin of Cooper pairs is zero ; the system obeys the full SRS. Consequently, the conductance is independent of the orientation of magnetization, i.e., $\sigma(\theta_{m},\phi_{m})=\sigma(\theta'_{m},\phi'_{m})$ for all values of $\theta_{m},\phi_{m},\theta'_{m}$ and $\phi'_{m}$. For the chiral $p$-wave superconductor with d-vector ${\bf{d}}=\Delta_{0}(k_{x}\pm ik_{y})\hat{\bf{z}}$ which is parallel to the crystallographic $c$-axis, the spin of Cooper pairs lie in the $ab$-plane. The system obeys the symmetry of the spin rotation about the $z$-axis. Consequently, the conductance is invariant under the magnetization rotation about the $z$-axis, i.e., $\sigma(\theta_{m},\phi_{m})=\sigma(\theta_{m},\phi'_{m})$ for all values of $\phi_{m}$ and $\phi'_{m}$. For the helical TS, the symmetry operation is the gauge-rotation $\mathcal{R}'(0,\pi)$ which `rotates' the direction of magnetization from $\hat{\bf{n}}(\theta_{m},\phi_{m})$ to $\hat{\bf{n}}(\theta_{m},\pi+\phi_{m})$ as given in table \ref{t2}. Thus, we obtain
\begin{eqnarray}
\sigma(\theta_{m},\phi_{m})=\sigma(\theta_{m},\pi+\phi_{m}).\label{con1}
\end{eqnarray}

The second situation which can lead to the invariance of conductance is the partial symmetries of the BdG Hamiltonian. The conductance $\sigma(\theta_{m},\phi_{m})$ is an average value of the angle-resolved conductance on $k_{y}$, which indicates that the junctions with $k_{y}$ and those with $-k_{y}$ correspond to the same conductance. For the $p_{xy}^{\pm}$-wave and the $p_{yx}^{\pm}$-wave TSs, the partial symmetry operations are $\mathcal{R}'(\pi,0)$, $\mathcal{R}(\pi,\pi)$ and $\mathcal{R}(\pi,0)$, $\mathcal{R}'(\pi,\pi)$, respectively, which transform the Hamiltonian $H(\bf{k})$ into $H(k_{x},-k_{y})$. They change the direction of magnetization from $\hat{\bf{n}}(\theta_{m},\phi_{m})$ to $\hat{\bf{n}}(\pi-\theta_{m},\pi-\phi_{m})$ and $\hat{\bf{n}}(\pi-\theta_{m},2\pi-\phi_{m})$. Consequently, we have
\begin{eqnarray}
\sigma(\theta_{m},\phi_{m})=\sigma(\pi-\theta_{m},\pi-\phi_{m}),\label{con2}\\
\sigma(\theta_{m},\phi_{m})=\sigma(\pi-\theta_{m},2\pi-\phi_{m}).\label{con3}
\end{eqnarray}
The jonit operations $\mathcal{R}'(0,\pi)\mathcal{R}'(\pi,0)$ and  the pure operation $\mathcal{R}(\pi,\pi)$ give the same symmetry relation of conductance due to the $U_{1}$ gauge symmetry satisfied by F.

The magnetization with $\theta_{m}=0$ or $\pi$ is a special case which means the magnetization is parallel to the spin quantization axis of the helical TS. In the coordinate of spin space in F, the TS are in purely equal-spin paring states. When an electron is injected from F, the normal reflected electron, the Andreev reflected hole and the transmitted quasiparticles all possess the same spin as that of the injected electron. As a result, the eight lines denoting the scattering processes presented in figure (\ref{f3}) will degenerate into four. The conductance in this situation is irrespective of the azimuthal angle $\phi_{m}$. The equations (\ref{con2}) and (\ref{con3}) will give $\sigma(\theta_{m}=0)=\sigma(\theta_{m}=\pi)$.

\begin{figure}[h!]
\begin{center}
\includegraphics[width=8cm]{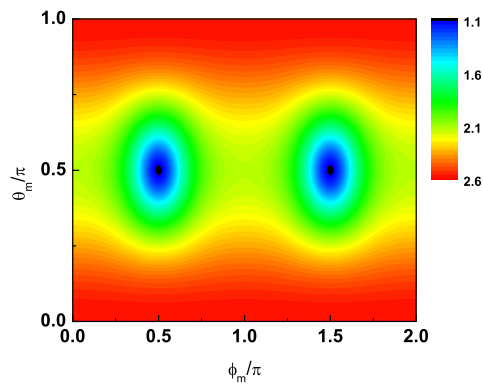}
\end{center}
\caption{(Color online) The conductance as a function of $\theta_{m}$ and $\phi_{m}$. The symmetry of the figure expresses the relations in equations (\ref{con1})-(\ref{con3}). The other parameters are taken as: $\frac{2mV_{0}}{\hbar^2k_{F}}=1$, $\frac{M}{E_{F}}=0.9$, $\frac{eV}{\Delta_{0}}=0.3$ and $T=0$ with $k_{F}$ the Fermi wavevector, $E_{F}$ the Fermi energy, $V$ the voltage and $T$ the temperature. The conductance is normalized by the normal value \cite{Cheng1}.}
\label{ff3}
\end{figure}
The obtained results in this part demonstrate that conductance, as an observable quantity, possesses a higher symmetry than the system itself. The SRS breaking of Hamiltonian favors strong anisotropy of conductance when the direction of magnetization is changed. However, the remaining partial symmetries of Hamiltonian help to keep invariance of conductance. In addition, the symmetry relations in equations (\ref{con1})-(\ref{con3}) have been proved in Ref. \cite{Cheng1} where the detailed derivation and numerical calculations are presented. Here, for the convenience of comparison between experiments and our results, we present in figure (\ref{ff3}) the conductance in the orientation space of magnetization, which can directly reflect the symmetry relations in equations (\ref{con1})-(\ref{con3}).

\section{Symmetries of Josephson current}
\subsection{Quasiclassical Green's function formalism}
\begin{figure}[h!]
\begin{center}
\includegraphics[width=7cm]{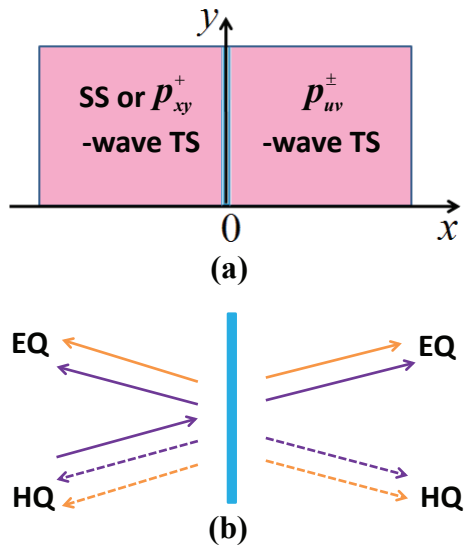}
\end{center}
\caption{(Color online) (a) Schematic illustration of the Josephson junctions considered in this paper. The $x$-axis is defined by the crystallographic $a$-axis. The magnetization in the ferromagnetic interface is assumed along $\hat{\bf{n}}(\theta_{m},\phi_{m})$. (b) The scattering process for an EQ injected from the left superconductor. The meaning of the lines is same as that in figure {\ref{f3}}. The spin for the injected EQ is assumed along the $z$-axis defined by the crystallographic $c$-axis; quasiparticles denoted by lines of the same color possess the same spin.}
\label{f4}
\end{figure}
The theory of quasiclassical Green's function is a powerful tool to study the Josephson effect in superconducting junctions, in which the Josephson current can be expressed by the retard Green's function. One method of obtaining the retard Green's function is to solve the Eilenberger equation \cite{Eilenberger}. The Riccati parametrization of the function and the diagrammatic representation of boundary conditions have been developed for the method \cite{Eschrig}. Another effective method of obtaining the retard Green's function is to construct it by composing wave functions derived from the BdG-equation $H({\bf{k}})\Psi=E({\bf{k}})\Psi$, which is innovated by McMillan \cite{McMillan} and extended to the case of anisotropic superconductor by Tanaka {\it et al}. \cite{Tanaka4} In this paper, the latter method is adopted.

We consider two-dimensional Josephson junctions with a ferromagnetic interface located $x=0$ and along the $y$-axis. The interface is described by the Hamiltonian in equation (\ref{HF}). For a given $k_{y}$ and the fixed spin, there are four types of quasiparticle injection process: the left or right injection of the electron-like quasiparticle and the hole-like quasiparticle. For each injection, there exist four scattering processes as shown in figure \ref{f4}. The wave function in superconductor is the superposition of the scattered quasiparticles. For the given $k_{y}$, there are four types of conjugated injection process with wavevector $-k_{y}$. The wave function in superconductor for the conjugated process can also be obtained from the superposition of the scattered quasiparticles. The retard Green's function is constructed from a linear combination of the product of wave functions and the transpose of the conjugated wave functions. The combination coefficients can be solved by using the boundary conditions at the ferromagnetic interface.

After an analytical continuation, the obtained Green's function $G(x,x',k_{y},i\omega_{n})$ is not only dependent on the spatial position, wavevector $k_{y}$ and the Matsubara frequency $\omega_{n}$ but also a function of $M$, $\hat{\bf{n}}(\theta_{m},\phi_{m})$, $\hat{V}_{0}$ and the superconducting phase difference $\Delta\phi$. The Josephson current for fixed $M$ and $V_{0}$ at the interface can be expressed as
\begin{eqnarray}
I_{J}(\theta_{m},\phi_{m},\Delta\phi)&=C_{1}\lim_{x'\rightarrow x}(\frac{\partial}{\partial x'}-\frac{\partial}{\partial x})\sum_{i\omega_{n},k_{y}}\mbox{Tr} [G(x,x',k_{y},\omega_{n})]\vert_{x=0}\\\label{IJ}
&=C_{2}\sum_{\omega_{n},k_{y}}F(i\omega_{n},k_{y}),
\end{eqnarray}
where $C_{1}$ and $C_{2}$ are constants which are irrelevant to the symmetries of the Josephson current. For simplicity, in the following we will assume the superconducting phase of the left superconductor is $\phi$ and that of the right is zero. Under this assumption, we have $I_{J}(\theta_{m},\phi_{m},\Delta\phi)=I_{J}(\theta_{m},\phi_{m},\phi)$. A concise discussion about how the symmetries of Hamiltonian affect the Josephson current is given in Appendix. In the following, we use equation (\ref{IJ}) as the starting point to analyze the properties of the current.

\subsection{SS$\vert$F$\vert$helical TS Josephson junctions}
We consider the Josephson junctions as shown in figure \ref{f4}(a). The SS with phase $\phi$ is located in the region with $x<0$ and the TS with zero phase in the region with $x>0$. They are described by the Hamiltonian in equations (\ref{Ham}) and (\ref{bdg}). The ferromagnetic interface is described by the Hamiltonian in equations (\ref{Ham}) and (\ref{HF}).

Firstly, we discuss the general results for the Josephson current of SS$\vert$F$\vert$helical TS junctions. For one thing, the time-reversal operation will change the superconducting phase of SS from $\phi$ to $-\phi$ according to equation (\ref{TSS}). The direction of the Josephson current is reversed accordingly, i.e., $I_{J}\rightarrow-I_{J}$. For another thing, the time-reversal `rotates' the magnetization in F from the direction $\hat{\bf{n}}(\theta_{m},\phi_{m})$ to $\hat{\bf{n}}(\pi-\theta_{m},\pi+\phi_{m})$. As a result, one obtains
\begin{eqnarray}
I_{J}(\theta_{m},\phi_{m},\phi)=-I_{J}(\pi-\theta_{m},\pi+\phi_{m},2\pi-\phi).\label{IJ1}
\end{eqnarray}

The $\mathcal{C}$ operator is another symmetry operation which can lead to a general result. The energy of quasiparticles becomes negative while the phase $\phi$ remains unchanged under the operation. Accordingly, the quantity $F(i\omega_{n},k_{y})$ will become $F(-i\omega_{n},k_{y})$ in equation (\ref{IJ}). However, the Josephson current $I_{J}$, as the sum of $F(i\omega_{n},k_{y})$, will keep its value. Actually, after the particle-hole transformation, the helical superconductor can be viewed as a new electron system with the normal dispersion $-\hat{\epsilon}(-{\bf{k}})$ and the energy gap matrix $\hat{\Delta}(-{\bf{k}})$. The SS can be viewed as a new electron system with the normal dispersion $-\hat{\epsilon}(-{\bf{k}})$ and the energy gap matrix $\hat{\Delta}(-{\bf{k}})e^{i\pi}$ according to equation (\ref{CSS}). The ferromagnetic interface becomes a new electron system with the normal dispersion $-\hat{\epsilon}(-{\bf{k}})$, the non-magnetic potential $-\hat{V}_{0}$ and the magnetization along $\hat{\bf{n}}(\pi-\theta_{m},\pi+\phi_{m})$ according to equation (\ref{CF}). As a result, we have
\begin{eqnarray}
I_{J}(\theta_{m},\phi_{m},\phi, V_{0})=I_{J}(\pi-\theta_{m},\pi+\phi_{m}, \pi+\phi,-V_{0}).\label{IJ2}
\end{eqnarray}
Since an actual interface possesses non-negative potential, i.e., $V_{0}\ge0$, only the case of $V_{0}=0$ for equation (\ref{IJ2}) makes sense. Then, we have
\begin{eqnarray}
I_{J}(\theta_{m},\phi_{m},\phi)=I_{J}(\pi-\theta_{m},\pi+\phi_{m}, \pi+\phi),\label{IJ3}
\end{eqnarray}
with $V_{0}=0$. Combing equations (\ref{IJ1}) and (\ref{IJ3}), we obtain
\begin{equation}
I_{J}(\theta_{m},\phi_{m},\phi)=-I_{J}(\theta_{m},\phi_{m},\pi-\phi),\label{IJ4}
\end{equation}
which is just the result brought by the chiral symmetry operation.

The operator $\mathcal{R}'(0,\pi)$ can also bring a general result. Under the operation, the helical TS is invariant as given in equation (\ref{gr1}) while the SS will acquire a $\pi$ phase as shown in table \ref{t1}. Meanwhile, the operation changes the direction of magnetization from $\hat{\bf{n}}(\theta_{m},\phi_{m})$ to $\hat{\bf{n}}(\theta_{m},\pi+\phi_{m})$. As a unitary transformation, $\mathcal{R}'(0,\pi)$ will not change the Josephson current. As a result, one obtains
\begin{eqnarray}
I_{J}(\theta_{m},\phi_{m},\phi)=I_{J}(\theta_{m},\pi+\phi_{m},\pi+\phi).\label{IJ5}
\end{eqnarray}

Secondly, we discuss other symmetry operations which will give different results for junctions with different pairing symmetries of the superconducting wave functions. Now, we take $\mathcal{R}'(\pi,0)$ as an example to analyze the symmetry of the Josephson current. Under the operation, the junctions with $k_{y}$ become those with $-k_{y}$ due to the partial symmetry of the $p_{xy}^{\pm}$-wave superconductor. Simultaneously, the $s$($d_{x^2-y^2}$)-wave SS acquires a phase of $\pi$ according to table {\ref{t1}}; the direction of magnetization turns into $\hat{\bf{n}}(\pi-\theta_{m},\pi-\phi_{m})$ according to table \ref{t2}. However, as an average quantity on $k_{y}$, the Josephson current in equation (\ref{IJ}) will not change its value. Thus, we have
\begin{eqnarray}
I_{J}(\theta_{m},\phi_{m},\phi)=I_{J}(\pi-\theta_{m},\pi-\phi_{m},\pi+\phi),\label{IJ6}
\end{eqnarray}
for the $s$($d_{x^2-y^2}$)-wave SS$\vert$F$\vert$$p_{xy}^{\pm}$-wave TS junctions. Similarly, from the operation $\mathcal{R}(\pi,\pi)$ we can derive
\begin{eqnarray}
I_{J}(\theta_{m},\phi_{m},\phi)=I_{J}(\pi-\theta_{m},2\pi-\phi_{m},\phi),\label{IJ7}
\end{eqnarray}
for the $s$($d_{x^2-y^2}$)-wave SS$\vert$F$\vert$$p_{xy}^{\pm}$-wave TS junctions. From the operations $\mathcal{R}'(\pi,0)$ and $\mathcal{R}(\pi,\pi)$, we can also derive
\begin{eqnarray}
I_{J}(\theta_{m},\phi_{m},\phi)=I_{J}(\pi-\theta_{m},\pi-\phi_{m},\phi),\label{IJ8}
\end{eqnarray}
and
\begin{eqnarray}
I_{J}(\theta_{m},\phi_{m},\phi)=I_{J}(\pi-\theta_{m},2\pi-\phi_{m},\pi+\phi),\label{IJ9}
\end{eqnarray}
respectively, for the $d_{xy}$-wave SS$\vert$F$\vert$$p_{xy}^{\pm}$-wave TS junctions.

For the SS$\vert$F$\vert$$p_{yx}^{\pm}$-wave TS junctions, there are similar symmetries of the Josephson current. The $s$($d_{x^2-y^2}$)-wave SS$\vert$F$\vert$$p_{yx}^{\pm}$-wave TS junctions satisfy equations (\ref{IJ8}) and (\ref{IJ9}) which originate from the operations $\mathcal{R}(\pi,0)$ and $\mathcal{R}'(\pi,\pi)$, respectively. The $d_{xy}$-wave SS$\vert$F$\vert$$p_{yx}^{\pm}$-wave TS junctions satisfy equations (\ref{IJ6}) and (\ref{IJ7}) which also originate from the operations $\mathcal{R}(\pi,0)$ and $\mathcal{R}'(\pi,\pi)$, respectively. From the results, we come to the following conclusions.

(a) The same spin-rotation or gauge-rotation operation can bring different symmetries of the Josephson current for the $s$($d_{x^2-y^2}$)-wave SS and the $d_{xy}$-wave SS when the gap function in the helical superconductor is fixed. The difference results from the partial symmetry obeyed by the helical superconductor and the different $k_{y}$-parities of gap functions in SS as discussed in part 2.1.2.

(b) The SS$\vert$$p_{uv}^{\pm}$-wave TS junctions can be classified into two types depending on the symmetries satisfied by their current. The $s$($d_{x^2-y^2}$)-wave SS$\vert$F$\vert$$p_{xy}^{\pm}$-wave TS junctions and the $d_{xy}$-wave SS$\vert$F$\vert$$p_{yx}^{\pm}$-wave TS junctions satisfy equations (\ref{IJ6}) and (\ref{IJ7}); the $s$($d_{x^2-y^2}$)-wave SS$\vert$F$\vert$$p_{yx}^{\pm}$-wave TS junctions and the $d_{xy}$-wave SS$\vert$F$\vert$$p_{xy}^{\pm}$-wave TS junctions satisfy equations (\ref{IJ8}) and (\ref{IJ9}). This classification is consistent with the numerical results in Ref. \cite{Cheng2}. Especially, the junctions of the same type possess the same selection rules of the lowest order Josephson current.

\begin{figure}[h!]
\begin{center}
\includegraphics[width=16cm]{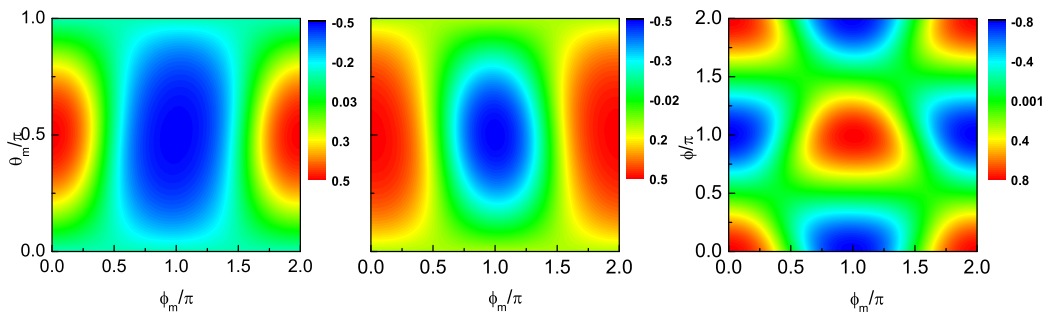}
\end{center}
\caption{(Color online) (a) The current for $\phi=0.3\pi$. (b) The current for $\phi=1.7\pi$. The symmetry of figure (a) expresses the equation (\ref{IJ7}). The connection between the two figures reflects the symmetry relations in equation (\ref{IJ1}). (c) The current for $\theta_{m}=0.3\pi$. The symmetry of figure (c) expresses the equation (\ref{IJ5}). The other parameters for all figures are taken as: $\frac{2mV_{0}}{\hbar^2k_{F}}=1$, $\frac{M}{E_{F}}=0.9$ and $T=0.3T_{C}$ with $T_{C}$ the critical temperature. The current has been expressed by the resistance in the normal state \cite{Cheng2}.}
\label{ff6}
\end{figure}
Finally, we show a brief explanation for the relation of the results in this paper to those in Ref. \cite{Cheng2}. The combination of equations here can bring equations in Ref. \cite{Cheng2}. For example, equations (\ref{IJ3}) and (\ref{IJ5}) will give $I_{J}(\theta_{m},\phi_{m},\phi)=I_{J}(\pi-\theta_{m},\phi_{m},\phi)$; equations (\ref{IJ3}) and (\ref{IJ6}) will give $I_{J}(\theta_{m},\phi_{m},\phi)=I_{J}(\theta_{m},2\pi-\phi_{m},\phi)$; equations(\ref{IJ3}) and (\ref{IJ9}) will give $I_{J}(\theta_{m},\phi_{m},\phi)=I_{J}(\theta_{m},\pi-\phi_{m},\phi)$. The three equalities are just the equations (8), (12) and (14) in Ref. \cite{Cheng2}, respectively. As pointed out in the reference, the three equations will not hold when the non-magnetic potential $V_{0}$ is non-zero (i.e., $Z\ne0$ there). This is because the equations are all based on equation (\ref{IJ3}) which holds only when the non-magnetic potential $V_{0}=0$. In order to compare our results with experiments easily, we present in figure (\ref{ff6}) the Josephson current for the SS$\vert$F$\vert p_{xy}^{+}$-wave TS junction in the orientation space of magnetization and in the space spanned by the azimuthal angle $\phi_{m}$ and the phase $\phi$, which can explicitly express the symmetry relations.

\subsection{Helical TS$\vert$F$\vert$Helical TS Josephson junctions}
We consider the purely helical Josephson junctions as shown in figure \ref{f4}(a). The $p_{xy}^{+}$-wave TS in the left side has the superconducting phase $\phi$. The right side can be the $p_{xy}^{-}$-wave or the $p_{yx}^{\pm}$-wave TS with the zero phase. The magnetization in the interface is assumed along $\hat{\bf{n}}(\theta_{m},\phi_{m})$.

Now ,we discuss the symmetries of the Josephson current. The time-reversal and the charge conjugation lead to
\begin{equation}
I_{J}(\theta_{m},\phi_{m},\phi)=-I_{J}(\pi-\theta_{m},\pi+\phi_{m},2\pi-\phi),\label{pIJ1}
\end{equation}
\begin{equation}
I_{J}(\theta_{m},\phi_{m},\phi,V_{0})=I_{J}(\pi-\theta_{m},\pi+\phi_{m},\phi, -V_{0}),\label{pIJ2}
\end{equation}
respectively, which are the common results for the purely helical junctions. The latter equality is different from equation (\ref{IJ2}). The helical superconductor will not acquire a phase $\pi$ under the charge conjugation.
\begin{figure}[h!]
\begin{center}
\includegraphics[width=14cm]{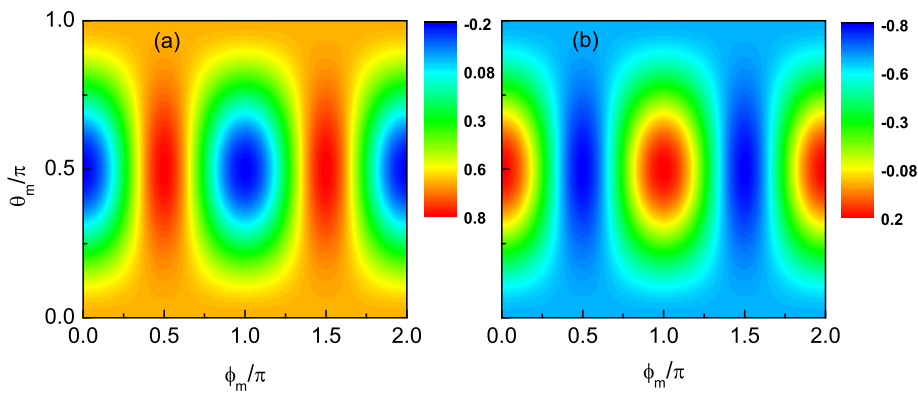}
\end{center}
\caption{(Color online) (a) The current for $\phi=0.3\pi$. (b) The current for $\phi=1.7\pi$. The symmetry of figure (a) expresses the equations (\ref{pIJ3}) and (\ref{pIJ4}). The connection between the two figures reflects the symmetry relations in equation (\ref{pIJ1}). The other parameters for the figures are taken as: $T=0.3T_{C}$, $\frac{M}{E_{F}}=0.9$ and $\frac{2mV_{0}}{\hbar^2k_{F}}=1$. The current has been expressed by the resistance in the normal state \cite{Cheng3}.}
\label{ff7}
\end{figure}

The rotation operations can bring different results for different junctions. For the $p_{xy}^{+}$-wave TS$\vert$F$\vert$$p_{xy}^{-}$-wave TS junction, $\mathcal{R}(0,\pi)$ and $\mathcal{R}(\pi,0)$ give
\begin{equation}
I_{J}(\theta_{m},\phi_{m},\phi)=I_{J}(\theta_{m},\pi+\phi_{m},\phi),\label{pIJ3}
\end{equation}
and
\begin{equation}
I_{J}(\theta_{m},\phi_{m},\phi)=I_{J}(\pi-\theta_{m},\pi-\phi_{m},\phi),\label{pIJ4}
\end{equation}
respectively, which are related to the transformations in equations (\ref{gr1}) and (\ref{gr2}). The result for $\mathcal{R}(\pi,\pi)$, as the joint operation of $\mathcal{R}(0,\pi)$ and $\mathcal{R}(\pi,0)$, is the combination of equations (\ref{pIJ3}) and (\ref{pIJ4}), i.e., $I_{J}(\theta_{m},\phi_{m},\phi)=I_{J}(\pi-\theta_{m},2\pi-\phi_{m},\phi)$. In figure (\ref{ff7}), we show the Josephson current for the $p_{xy}^{+}$-wave TS$\vert$F$\vert$$p_{xy}^{-}$-wave TS junction in the orientation space of magnetization. It may be useful in comparing our results with experiments. For simplicity, we will not give numerical results for other junctions.

For the $p_{xy}^{+}$-wave TS$\vert$F$\vert$$p_{yx}^{\pm}$-wave TS junctions, $\mathcal{R}(0,\pi)$ and $\mathcal{R}(\pi,0)$ give
\begin{equation}
I_{J}(\theta_{m},\phi_{m},\phi)=I_{J}(\theta_{m},\pi+\phi_{m},\phi),\label{pIJ5}
\end{equation}
and
\begin{equation}
I_{J}(\theta_{m},\phi_{m},\phi)=I_{J}(\pi-\theta_{m},\pi-\phi_{m},\pi+\phi),\label{pIJ6}
\end{equation}
respectively, which are related to the transformations in equations (\ref{gr1}), (\ref{gr2}) and (\ref{pr2}). The result for $\mathcal{R}(\pi,\pi)$, as the joint operation of $\mathcal{R}(0,\pi)$ and $\mathcal{R}(\pi,0)$, is the combination of equations (\ref{pIJ5}) and (\ref{pIJ6}), i.e., $I_{J}(\theta_{m},\phi_{m},\phi)=I_{J}(\pi-\theta_{m},2\pi-\phi_{m},\pi+\phi)$. There is a $\pi$-phase difference between equation (\ref{pIJ4}) and (\ref{pIJ6}) since the $p_{xy}^{+}$-wave TS is partially symmetric under $\mathcal{R}'(\pi,0)$ while $\mathcal{R}(\pi,0)$ will bring an extra phase $\pi$ to the superconductor.

The helical Josephson junctions can host rich ground states which can be classified into $0$ phase, $\pi$ phase, $0+\pi$ phase, $\varphi_{0}$ phase and $\varphi$ phase \cite{Cheng3}. The phase transition can be tuned by rotating the magnetization in
the ferromagnetic interface. The phase diagram formed in the orientation space of magnetization is an important aspect of researches on Josephson junctions which also possesses some symmetries as results of symmetries of the current.

For the $p_{xy}^{+}$-wave TS$\vert$F$\vert$$p_{xy}^{-}$-wave TS junction, the diagram in the orientation space $0\le\theta_{m}<\pi$ and $0\le\phi<2\pi$ is symmetric about both $\theta_{m}=\pi/2$ and $\phi_{m}=\pi$. It is the result of equations (\ref{pIJ3}), (\ref{pIJ4}) and
\begin{equation}
I_{J}(\theta_{m},\phi_{m},\phi)=I_{J}(\theta_{m},\pi-\phi_{m},\phi).\label{pIJ7}
\end{equation}
Equation (\ref{pIJ7}) originates from the close relation between $p_{xy}^{\pm}$-wave TSs. When one changes the $y$-axis into $-y$ in the three dimensional coordinate space, the TSs exchange each other; the direction of magnetization in the new coordinate system becomes $\hat{\bf{n}}(\theta_{m},2\pi-\phi_{m},\phi)$ accordingly. The relation can be reflected directly from the interaction of helical superconductivity and ferromagnetism which can be seen in part 5.

For the $p_{xy}^{+}$-wave TS$\vert$F$\vert$$p_{yx}^{\pm}$-wave TS junctions, the phase diagram for $V_{0}=0$ is symmetric about $\theta_{m}=\pi/2$ and invariant under the translation $\phi_{m}\rightarrow\phi_{m}+\pi$. It is the result of equations (\ref{pIJ5}) and (\ref{pIJ2}) with $V_{0}=0$. In addition, we have another equality,
\begin{equation}
I_{J}(\theta_{m},\phi_{m},\phi)=I_{J}(\theta_{m},\pi/2-\phi_{m},\phi),\label{pIJ8},
\end{equation}
which originates from the connection between the TSs. For example, for the $p_{xy}^{+}$-wave TS$\vert$F$\vert$$p_{yx}^{+}$-wave TS junction, when one changes the $x$-axis and the $y$-axis into $y$ and $x$ respectively, the TSs exchange each other; the direction of the magnetization becomes $\hat{\bf{n}}(\theta,\pi/2-\phi_{m},\phi)$ in the new coordinate space. The connection can also be reflected directly from the interaction of helical superconductivity and ferromagnetism which can be seen in part 5.

As discussed in part 3.2 for the conductance, the magnetization with $\theta_{m}=0$ or $\pi$ is also a special case for the Josephson current. The purely equal-spin paring will simplify the scattering process in figure (\ref{f4}). The current will become irrespective of the azimuthal angle $\phi_{m}$. For the SS$\vert$F$\vert$TS and TS$\vert$F$\vert$TS junctions, we will have $I_{J}(\theta_{m}=0,\phi)=-I_{J}(\theta_{m}=\pi,2\pi-\phi)$ from equations (\ref{IJ1}) and (\ref{pIJ1}). This relation can also be found in the Josephson junction formed along the edges of quantum spin-Hall insulators with a Zeeman field along the spin quantization \cite{Dolcini2}.

Another special case is the perpendicular magnetization with $\theta_{m}=\pi/2$ which is important for helical superconductor realized in a topological insulator with the $s$-wave superconductor \cite{Fu2,Crepin}. The symmetries of the transport quantities in these references are different from those in this paper. For example, in reference \cite{Crepin}, the transmission probability of the normal-metal-superconductor junction with a single ferromagnetic barrier is independent of the direction of the in-plane magnetization. For the normal-metal-superconductor junction with double barriers, the probability is dependent on the relative angle of the in-plane magnetization in the barriers. However, in our junctions with $\theta_{m}=\pi/2$, the conductance and the current is strongly dependent on the azimuthal angle of magnetization due to the symmetry breaking of the spin-rotation.

Finally, we discuss the Josephson current in the non-magnetic case, i.e., $M=0$. In this case, we have $I_{J}(\phi)=I_{J}(\pi+\phi)$ for the SS$\vert$TS junctions which means the Josephson current is $\pi$-periodic. This result is consistent with that for the junction between singlet and triplet superconductors in the static thermodynamic limit \cite{Kwon}. For the $p_{xy}^{+}$-wave TS$\vert$$p_{yx}^{\pm}$-wave TS junctions, we also have $I_{J}(\phi)=I_{J}(\pi+\phi)$ with $\pi$ periodicity which also indicates the absence of the $\sin{\phi}$-type current. This result is also consistent with that for the triplet junction in which the two $d$-vectors are perpendicular \cite{Kwon}. For the helical states with $k$-dependent $d$-vectors in this paper, the perpendicular vectors possess the relation $\langle {\bf{d}}_{xy}^{+}\cdot{\bf{d}}_{yx}^{\pm}\rangle_{k_{y}}=0$ in which $\langle\cdot\cdot\cdot\rangle_{k_{y}}$ denotes the average over the momentum parallel to the interface \cite{Cheng3}.

\section{Symmetries of free energy}
In the above sections we have analyzed the symmetries of the current in Josephson junctions from the viewpoint of symmetries of Hamiltonian. Actually, the current-phase relation has a more direct relationship with free energy of junctions, which is derivative of the free energy with respect to $\phi$. The free energy of the Josephson junctions have been constructed in Ref. \cite{Cheng2} and \cite{Cheng3} on the base of numerical results, which directly reflects the interaction of helical superconductivity and ferromagnetism. In the following sections we will show that symmetries of the current derived from Hamiltonian are consistent with those derived from the free energy.

\subsection{SS$\vert$F$\vert$helical TS Josephson junctions}
The free energy for the $s$($d_{x^2-y^2})$-wave SS$\vert$F$\vert$$p_{xy}^{\pm}$-wave TS and the $d_{xy}$-wave SS$\vert$F$\vert$$p_{yx}^{\pm}$-wave TS junctions, denoted by $F_{ST1}$, is given by
\begin{equation}
F_{ST1}\propto\sin{\theta_{m}}\cos{\phi_{m}}\sin{\phi},\label{fe1}
\end{equation}
for $V_{0}=0$ and
\begin{equation}
F_{ST1}\propto\sin{\theta_{m}}\cos{\phi_{m}}\sin{\phi}~~~\&~~~\sin{\theta_{m}}\cos{\theta_{m}}\sin{\phi_{m}}\cos{\phi},\label{fe2}
\end{equation}
for $V_{0}\ne0$. The proportional coefficient before each term in the expressions is generally a complex function of parameters such as $M$, $V_{0}$ and $\Delta_{0}$, which is irrespective of the symmetries of the free energy. We have omitted them for simplicity. The symbol $\&$ suggests that there are two terms contributed to the current when $V_{0}\ne0$. The Josephson current is given accordingly by
\begin{equation}
I_{J}\propto\sin{\theta_{m}}\cos{\phi_{m}}\cos{\phi},\label{fec1}
\end{equation}
for $V_{0}=0$ and
\begin{equation}
I_{J}\propto\sin{\theta_{m}}\cos{\phi_{m}}\cos{\phi}~~~\&~~~\sin{\theta_{m}}\cos{\theta_{m}}\sin{\phi_{m}}\sin{\phi},\label{fec2}
\end{equation}
for $V_{0}\ne0$. It is easy to verify that the current for $V_{0}=0$ satisfies equations (\ref{IJ1}) and (\ref{IJ3})-(\ref{IJ7}). However, equations (\ref{IJ3}) and (\ref{IJ4}) do not hold for the current with $V_{0}\ne0$ due to the presence of the second term in equation (\ref{fec2}), which is consistent with the true condition of equations (\ref{IJ3}) and (\ref{IJ4}).

The free energy for the $s$($d_{x^2-y^2})$-wave SS$\vert$F$\vert$$p_{yx}^{\pm}$-wave TS and the $d_{xy}$-wave SS$\vert$F$\vert$$p_{xy}^{\pm}$-wave TS junctions, denoted by $F_{ST2}$, is given by
\begin{equation}
F_{ST2}\propto\sin{\theta_{m}}\sin{\phi_{m}}\sin{\phi},\label{fe3}
\end{equation}
for $V_{0}=0$ and
\begin{equation}
F_{ST2}\propto\sin{\theta_{m}}\sin{\phi_{m}}\sin{\phi}~~~\&~~~\sin{\theta_{m}}\cos{\theta_{m}}\cos{\phi_{m}}\cos{\phi},\label{fe4}
\end{equation}
for $V_{0}\ne0$. The Josephson current is accordingly given by
\begin{equation}
I_{J}\propto\sin{\theta_{m}}\sin{\phi_{m}}\cos{\phi},\label{fec3}
\end{equation}
for $V_{0}=0$ and
\begin{equation}
I_{J}\propto\sin{\theta_{m}}\sin{\phi_{m}}\cos{\phi}~~~\&~~~\sin{\theta_{m}}\cos{\theta_{m}}\cos{\phi_{m}}\sin{\phi},\label{fec4}
\end{equation}
for $V_{0}\ne0$. It is easy to verify that the current for $V_{0}=0$ satisfies equations (\ref{IJ1}), (\ref{IJ3})-(\ref{IJ5}), (\ref{IJ8}) and (\ref{IJ9}). However, equations (\ref{IJ3}) and (\ref{IJ4}) do not hold for the current with $V_{0}\ne0$ due to the presence of the second term in equation (\ref{fec4}), which is consistent with the true condition of equations (\ref{IJ3}) and (\ref{IJ4}).

\subsection{Helical TS$\vert$F$\vert$helical TS Josephson junctions}
The free energy for the $p_{xy}^{+}$-wave TS$\vert$F$\vert$$p_{xy}^{-}$-wave TS junction, denoted by $F_{TT1}$, is given by
\begin{equation}
F_{TT1}\propto\cos^{2}\theta_{m}\cos{\phi}~~~\&~~~\sin^{2}\theta_{m}\cos2\phi_{m}\cos{\phi}.\label{pfe1}
\end{equation}
The Josephson current is accordingly given as
\begin{equation}
I_{J}\propto\cos^{2}\theta_{m}\sin{\phi}~~~\&~~~\sin^{2}\theta_{m}\cos2\phi_{m}\sin{\phi}.\label{pfec1}
\end{equation}
It can be easily demonstrated that the current satisfies equations (\ref{pIJ1})-(\ref{pIJ4}). In addition, equation (\ref{pIJ7}) can be derived directly from equation (\ref{pfec1}) which is not obeyed by the SS$\vert$F$\vert$helical $p$-wave TS junctions due to the absence of the close relation discussed in part 4.3.

The free energy of the $p_{xy}^{+}$-wave TS$\vert$F$\vert$$p_{yx}^{\pm}$-wave TS junctions, denoted by $F_{TT2}$, is given by
\begin{equation}
F_{TT2}\propto\sin^{2}{\theta_{m}}\sin{2\phi_{m}}\cos{\phi},\label{pfe2}
\end{equation}
for $V_{0}=0$ and
\begin{equation}
F_{TT2}\propto\sin^{2}{\theta_{m}}\sin{2\phi_{m}}\cos{\phi}~~~\&~~~\cos{\theta_{m}}\sin{\phi},\label{pfe3}
\end{equation}
for $V_{0}\ne0$. The Josephson current is accordingly given as
\begin{equation}
I_{J}\propto\sin^{2}{\theta_{m}}\sin{2\phi_{m}}\sin{\phi},\label{pfec2}
\end{equation}
for $V_{0}=0$ and
\begin{equation}
I_{J}\propto\sin^{2}{\theta_{m}}\sin{2\phi_{m}}\sin{\phi}~~~\&~~~\cos{\theta_{m}}\cos{\phi},\label{pfec3}
\end{equation}
for $V_{0}\ne0$. It can be demonstrated that the current satisfies equations (\ref{pIJ1}), (\ref{pIJ2}), (\ref{pIJ5}) and (\ref{pIJ6}). In addition, equations (\ref{pIJ7}) and (\ref{pIJ8}) can be directly derived from equations (\ref{pfec2}) and (\ref{pfec3}), which are not satisfied by the SS$\vert$F$\vert$helical $p$-wave TS junctions due to the absence of the connection discussed in part 4.3.

\section{Conclusions}
We establish linkages between symmetries of the Hamiltonian for topological superconducting systems and the invariance of transport quantities for the corresponding junctions. As observables, transport quantities exhibit higher symmetries than the systems themselves. We reveal the important role that partial symmetries play in the invariance of conductance and Josephson current. Our analysis explain the numerical results reported recently which include the selection rules of the lowest order current and the rich phase diagrams in Josephson junctions. The analysis method not only provides a profound understanding of topological junctions but also helps give useful information about transport quantities before carrying out complex numerical calculations. In addition, the symmetry analysis of transport properties in this paper is general and applicable for other topological junctions such as the chiral ones. However, the present studies do not include the $4\pi$-periodic Josephson effect caused by transmission of unpaired electrons \cite{Beenakker3}. The novel effect is related to the fermion-parity anomaly in the superconducting ground state.

\section*{ACKNOWLEDGMENTS}
This work is supported in part by the National Natural Science Foundation of China (Grant Nos. 11447175, 11547035, 11505100 and 61572270), the Natural Science Foundation of Shandong Province (Grant No. ZR2015AQ007) and the Qingdao Science and Technology Program (Grant No. 14-2-4-110-JCH).

\section*{Appendix}
\setcounter{equation}{0}
\renewcommand{\theequation}{A.\arabic{equation}}
Taking the F$\vert$$p_{xy}^{+}$-wave TS junction in figure \ref{f3}(a) as an example, we show in detail how the symmetries of conductance are obtained from the symmetries of the Hamiltonian for the helical superconductor.

The BdG Hamiltonian for the ferromagnetic region $(x<0)$ is
\begin{eqnarray}\label{A1}
H_{F}({\bf{k}})=\left(
\begin{array}{cc}
\hat{\epsilon}({\bf{k}})-{\bf{M}}\cdot\hat{\sigma}&0\\
0&-\hat{\epsilon}^{*}({-\bf{k}})+{\bf{M}}\cdot\hat{\sigma}^{*}
\end{array}\right),
\end{eqnarray}
with $\hat{\epsilon}({\bf{k}})=(\frac{\hbar^2k^2}{2m}-E_{F})\hat{1}$ and ${\bf{M}}=M\hat{{\bf{n}}}(\theta_{m},\phi_{m})$ specified by the direction $\hat{\bf{n}}=(\sin\theta_{m}\cos{\phi_{m}}, \sin{\theta_{m}}\sin{\phi_{m}}, \cos{\theta_{m}})$ with the polar angle $\theta_{m}$ and the azimuthal angle $\phi_{m}$. For the spin of electrons, we chose the direction of ${\bf{M}}$ as the quantization axis. Let us consider an electron with majority spin (spin-up) is injected from F. The wave functions can be given by
\begin{eqnarray}
\Psi^{F}(x<0)=\psi^{F}_{e\uparrow}e^{ik_{e\uparrow}x}+b_{\uparrow\uparrow}\psi_{e\uparrow}^{F}e^{-ik_{e\uparrow}x}
+b_{\uparrow\downarrow}\psi_{e\downarrow}^{F}e^{-ik_{e\downarrow}x}\\\nonumber
+a_{\uparrow\uparrow}\psi^{F}_{h\uparrow}e^{ik_{h\uparrow}x}+a_{\uparrow\downarrow}\psi^{F}_{h\downarrow}e^{ik_{h\downarrow}x},
\end{eqnarray}
where $\psi^{F}_{e\uparrow}=(\chi_{1},\chi_{2},0,0)^{T}$, $\psi^{F}_{e\downarrow}=(-\chi_{2}^{*},\chi_{1},0,0)^{T}$, $\psi^{F}_{h\uparrow}=(0,0,\chi_{1},\chi_{2}^{*})^{T}$ and $\psi^{F}_{h\downarrow}=(0,0,-\chi_{2},\chi_{1})^{T}$ with $\chi_{1}=\cos{\frac{\theta_{m}}{2}}$ and $\chi_{2}=\sin{\frac{\theta_{m}}{2}}e^{i\phi_{m}}$; the wavevectors $k_{e\uparrow}=k_{h\uparrow}=\sqrt{\frac{2m}{\hbar}(E_{F}+M-k_{y}^2)}$ and $k_{e\downarrow}=k_{h\downarrow}=\sqrt{\frac{2m}{\hbar}(E_{F}-M-k_{y}^2)}$. The coefficients $b_{\uparrow\uparrow}(b_{\uparrow\downarrow})$ and $a_{\uparrow\uparrow}(a_{\uparrow\downarrow})$ represent the normal reflection to majority (minority) spin subband and the Andreev reflection to majority (minority) spin suband, respectively, which correspond to the scattering process shown in figure $\ref{f3}$(b).

The BdG Hamiltonian for the superconducting region $(x>0)$ is
\begin{eqnarray}\label{A4}
H({\bf{k}})=\left(
\begin{array}{cc}
\hat{\epsilon}({\bf{k}})&\hat{\Delta}({\bf{k}})\\
-\hat{\Delta}^{*}({-\bf{k}})&-\hat{\epsilon}^{*}(-{\bf{k}})
\end{array}\right),
\end{eqnarray}
where $\hat{\Delta}({\bf{k}})=({\bf{d}}_{xy}^{+}({\bf{k}})\cdot\hat{\sigma})i\hat{\sigma}_{2}$ with ${\bf{d}}_{xy}^{+}({\bf{k}})=\Delta_{0}(k_{x}\hat{x}+k_{y}\hat{y})$. The wave function is given by
\begin{eqnarray}
\Psi_{S}(x>0)=c_{\uparrow\uparrow}\psi_{e\uparrow}^{S}e^{ik_{x}x}+c_{\uparrow\downarrow}\psi_{e\downarrow}^{S}e^{ik_{x}x}
+d_{\uparrow\uparrow}\psi_{h\uparrow}^{S}e^{-ik_{x}x}+d_{\uparrow\downarrow}\psi_{h\downarrow}^{S}e^{-ik_{x}x},
\end{eqnarray}
where $\psi_{e\uparrow}^{S}=(u,0,-v\eta(k_{y}),0)^{T}$, $\psi_{e\uparrow}^{S}=(0,u,0,v\eta^{*}(k_{y}))^{T}$,
$\psi_{e\uparrow}^{S}=(-v\eta^{*}(-k_{y}),0,u,0)^{T}$ and $\psi_{e\uparrow}^{S}=(0,v\eta(-k_{y}),0,u)^{T}$ with
$u(v)=\sqrt{\frac{E+(-)\Omega}{2E}}$ and $\Omega=\sqrt{E^2-\Delta_{0}^2}$; the phase factor $\eta(k_{y})=\frac{k_{x}+ik_{y}}{k_{F}}$ with $k_{x}=\sqrt{k_{F}^2-k_{y}^2}$. The coefficients $c_{\uparrow\uparrow}, c_{\uparrow\downarrow}, d_{\uparrow\uparrow}$ and $d_{\uparrow\downarrow}$ represent the transmission of electron-like quasiparticle and hole-like quasiparticle, respectively.

All the coefficients in the wave functions can be determined under the boundary conditions:
\begin{eqnarray}
\Psi_{F}(x=0^{-})=\Psi_{S}(x=0^{+}),\label{A5}\\
\Psi^{'}_{S}(x=0^{+})-\Psi^{'}_{F}(x=0^{-})=\frac{2mV_{0}}{\hbar^2}\Psi_{F}(x=0).\label{A6}
\end{eqnarray}
Obviously, the obtained coefficients will be functions of $\theta_{m}$, $\phi_{m}$, $k_{y}$ and $V_{0}$. The coefficients for an injection electron with the minority spin can be obtained in a similar way. According to the Blonder-Tinkham-Klapwijk formalism, the conductance for a given $V_{0}$ can be written as $\sigma(\theta_{m},\phi_{m})=C_{0}\sum_{k_{y}}\sigma'(\theta_{m},\phi_{m},k_{y})$ where
\begin{eqnarray}
\sigma'(\theta_{m},\phi_{m},k_{y})=\sigma_{\uparrow}+\sigma_{\downarrow},\label{A7}
\end{eqnarray}
\begin{eqnarray}
\label{A8}
\sigma_{\uparrow}=\frac{1+X}{2}(1+\vert a_{\uparrow\uparrow}(\theta_{m},\phi_{m},k_{y})\vert^2+\frac{k_{h\downarrow}}{k_{e\uparrow}}\vert a_{\uparrow\downarrow}(\theta_{m},\phi_{m},k_{y})\vert^2\\\nonumber
-\vert b_{\uparrow\uparrow}(\theta_{m},\phi_{m},k_{y})\vert^2-\frac{k_{e\downarrow}}{k_{e\uparrow}}\vert b_{\uparrow\downarrow}(\theta_{m},\phi_{m},k_{y})\vert^2),
\end{eqnarray}
and
\begin{eqnarray}
\label{A9}
\sigma_{\downarrow}=\frac{1-X}{2}(1+\frac{k_{h\uparrow}}{k_{e\downarrow}}\vert a_{\downarrow\uparrow}(\theta_{m},\phi_{m},k_{y})\vert^2+\vert a_{\downarrow\downarrow}(\theta_{m},\phi_{m},k_{y})\vert^2\\\nonumber
-\frac{k_{e\uparrow}}{k_{e\downarrow}}\vert b_{\uparrow\downarrow}(\theta_{m},\phi_{m},k_{y})\vert^2-\vert b_{\downarrow\downarrow}(\theta_{m},\phi_{m},k_{y})\vert^2),
\end{eqnarray}
with $X=\frac{M}{E_{F}}$.

In order to derive the symmetry relations in equations (\ref{con1})-(\ref{con3}), we firstly transform the Hamiltonian for superconductor in equation (\ref{A4}) with the gauge-rotation operation $\mathcal{R}'(0,\pi)$. The transformation does not change $H({\bf{k}})$; the wave function in the superconducting region remain unchanged. The same operation can change the Hamiltonian $H_{F}({\bf{k}})$ with $\hat{\bf{n}}(\theta_{m},\phi_{m})$ in equation (\ref{A1}) into $H_{F}({\bf{k}})$ with $\hat{\bf{n}}(\theta_{m},\pi+\phi_{m})$; the wave function in the ferromagnetic region can correspondingly be obtained through the substitution of $\pi+\phi_{m}$ for $\phi_{m}$ in $\Psi_{F}$. Under the boundary conditions in equations (\ref{A5}) and (\ref{A6}), the coefficients for the transformed system can be determined, which are functions of $\theta_{m}$, $\pi+\phi_{m}$ and $k_{y}$ for a given $V_{0}$. The conductance can be obtained by the substitution of $\pi+\phi_{m}$ for $\phi_{m}$ in equations (\ref{A7})-(\ref{A9}), i.e., $\sigma(\theta_{m},\pi+\phi_{m})=C_{0}\sum_{k_{y}}\sigma'(\theta_{m},\pi+\phi_{m},k_{y})$. As a unitary operator, $\mathcal{R}'(0,\pi)$ will keep the conductance invariant under the transformation. Then we have the relation $\sigma(\theta_{m},\phi_{m})=\sigma(\theta_{m},\pi+\phi_{m})$.

Secondly, we consider the operations $\mathcal{R}'(\pi,0)$ and $\mathcal{R}(\pi,\pi)$. They not only transform the direction $\hat{\bf{n}}(\theta_{m},\phi_{m})$ in $H_{F}({\bf{k}})$ into $\hat{\bf{n}}(\pi-\theta_{m},\pi-\phi_{m})$ and $\hat{\bf{n}}(\pi-\theta_{m},2\pi-\phi_{m})$, respectively, but also change the superconducting Hamiltonian $H({\bf{k}})$ into $H(k_{x},-k_{y})$. The wave functions for the transformed Hamiltonian in the ferromagnetic region can be obtained by the similar substitutions for $\theta_{m}$ and $\phi_{m}$ in $\Psi_{F}$ as discussed; the wave function in the superconducting region can be obtained by replacing $k_{y}$ with $-k_{y}$ in $\Psi_{S}$. Under the boundary conditions, the coefficients and thus the conductance for the transformed system can be determined, i.e., $\sigma(\pi-\theta_{m},\pi-\phi_{m})=C_{0}\sum_{k_{y}}\sigma'(\pi-\theta_{m},\pi-\phi_{m},-k_{y})
=C_{0}\sum_{k_{y}}\sigma'(\pi-\theta_{m},\pi-\phi_{m},k_{y})$ and $\sigma(\pi-\theta_{m},2\pi-\phi_{m})=C_{0}\sum_{k_{y}}\sigma'(\pi-\theta_{m},2\pi-\phi_{m},-k_{y})
=C_{0}\sum_{k_{y}}\sigma'(\pi-\theta_{m},2\pi-\phi_{m},k_{y})$. Then we have the relations in equations (\ref{con2}) and (\ref{con3}) due to the unitarity of $\mathcal{R}'(\pi,0)$ and $\mathcal{R}(\pi,\pi)$. For other F$\vert$ helical TS junctions, one can derive the symmetry relations in a similar way.

Finally, we discuss how to derive symmetries of the Josephson current from the symmetries satisfied by Hamiltonian. In order to express the current, we first need to obtain the retard Green's function which can be constructed with the scattering wave function in superconductors \cite{McMillan,Tanaka4}. The wave function and the scattering coefficients can be obtained by solving the BdG equation under boundary conditions. The solving process is the same as that for the conductance case. Then, the changes of Hamiltonian under the symmetry transformations will enter into the wave function and the coefficients. Actually, the finial expression of the current can be written as the combination of the Andreev reflection coefficients \cite{Furusaki2}, i.e., $F(i\omega_{n},k_{y})$ in equation (\ref{IJ}) is proportional to $a(\theta_{m},\phi_{m},i\omega_{n},k_{y},\phi)$. The effects of symmetry transformations on the Andreev coefficient have been discussed carefully. The symmetries of the current can be derived in a similar way.


\section*{References}
\begin{thebibliography}{99}
\bibitem{Hasan} Hasan M Z and Kane C L 2010 {\it Rev. Mod. Phys.} \textbf{82} 3045.
\bibitem{Qi} Qi X L and Zhang S C 2011 {\it Rev. Mod. Phys.} \textbf{83} 1057.
\bibitem{Tanaka} Tanaka Y, Sato M and Nagaosa N 2012 {\it J. Phys. Soc. Jpn.} \textbf{81} 011013.
\bibitem{Ando} Ando Y and Fu L 2015 {\it Annu. Rev. Condens. Matter. Phys.} \textbf{6} 361.
\bibitem{Chiu} Chiu C K, Teo J C Y, Schnyder A P and Ryu S 2016 {\it Rev. Mod. Phys.} \textbf{88} 035005.
\bibitem{Sato} Sato M and Ando Y {\it arXiv:1608.03395 [cond-mat.supr-con]}.
\bibitem{Kane} Kane C L and Mele E J 2005 {\it Phys. Rev. Lett.} \textbf{95} 146802.
\bibitem{Sato2} Sato M, Tanaka Y, Yada K and Yokoyama T 2011 {\it Phys. Rev. B} \textbf{83} 224511.
\bibitem{Sato3} Sato M 2006 {\it Phys. Rev. B} \textbf{73} 214502.
\bibitem{Kimme} Kimme L, Hyart T and Rosenow B 2015 {\it Phys. Rev. B} \textbf{91} 220501.
\bibitem{Wakatsuki} Wakatsuki R, Ezawa M, Tanaka Y and Nagaosa N 2014 {\it Phys. Rev. B} \textbf{90} 014505.
\bibitem{Ezawa} Ezawa M, Tanaka Y and Nagaosa N 2013 {\it Sci. Rep.} \textbf{3} 2790.
\bibitem{Ryu} Ryu S, Schnyder A P, Furusaki and Ludwig A W W 2010 {\it New J. Phys.} \textbf{12} 065010.
\bibitem{Schnyder} Schnyder A P, Ryu S, Furusaki A and Ludwig A W W 2009 {\it AIP Conf. Proc.} \textbf{1134} 10.
\bibitem{Schnyder2} Schnyder A P, Ryu S, Furusaki A and Ludwig A W W 2008 {\it Phys. Rev. B} \textbf{78} 195125.
\bibitem{Kane2} Kane C L and Mele E J 2005 {\it Phys. Rev. Lett.} \textbf{95} 146802.
\bibitem{Bernevig} Bernevig B A and Zhang S C 2006 {\it Phys. Rev. Lett.} \textbf{96} 106802.
\bibitem{Mackenzie} Mackenzie A P and Maeno Y 2003 {\it Rev. Mod. Phys.} \textbf{75} 657.
\bibitem{Maeno} Maeno Y, Kittaka S, Nomura T, Yonezawa S and Ishida K 2012 {\it J. Phys. Soc. Jpn.} \textbf{81} 011009.
\bibitem{Zhang} Zhang J, Lorscher C, Gu Q and Klemm R A 2014 {\it J. Phys.: Condens. Matter} \textbf{26}
\bibitem{Bauer} Bauer E, Hilscher G, Michor H, Paul Ch, Scheidt E W, Gribanov A, Seropegin Yu, Noel H, Sigrist M and Rogl P 2004 {\it Phys. Rev. Lett.} \textbf{92} 027003.
\bibitem{Mizushima} Mizushima T, Tsutsumi Y, Kawakami T and Sato M 2016 {\it J. Phys. Soc. Jpn} \textbf{85} 022001.
\bibitem{Furusaki} Furusaki A, Matsumoto M and Sigrist M 2001 {\it Phys. Rev. B} \textbf{64} 054514.
\bibitem{Fu} Fu L and Kane C L 2009 {\it Phys. Rev. Lett.} \textbf{102} 216403.
\bibitem{Zuo} Zuo Z W, Li H J, Li L B, Sheng L, Shen R and Xing D Y 2016 {\it Europhys. Lett.} \textbf{114} 27001.
\bibitem{Tanaka2} Tanaka Y, Yokoyama T and Nagaosa N 2009 {\it Phys. Rev. Lett.} \textbf{103} 107002.
\bibitem{Lutchyn} Lutchyn R M, Sau J D and Sarma S D 2010 {\it Phys. Rev. Lett.} \textbf{105} 077001.
\bibitem{Linder} Linder J, Tanaka Y, Yokoyama T, Sudbo A and Nagaosa N 2010 {\it Phys. Rev. Lett.} \textbf{104} 067001.
\bibitem{Asano} Asano Y and Tanaka Y 2013 {\it Phys. Rev. B} \textbf{87} 104513.
\bibitem{Beenakker1} Beenakker C W J 2015 {\it Rev. Mod. Phys.} \textbf{87} 1037.
\bibitem{Beenakker2} Beenakker C W J 2013 {\it Annu. Rev. Condens. Matter. Phys.} \textbf{4} 113.
\bibitem{Kashiwaya} Kashiwaya S and Tanaka Y 2000 {\it Rep. Prog. Phys.} \textbf{63} 1641.
\bibitem{Lu} Lu C K and Yip S K 2009 {\it Phys. Rev. B} \textbf{80} 024504.
\bibitem{Tanaka3} Tanaka Y and Kashiwaya S 1995 {\it Phys. Rev. Lett.} \textbf{74} 3451.
\bibitem{Hirai} Hirai T, Tanaka Y, Yoshida N, Asano Y, Inoue J and Kashiwaya S 2003 {\it Phys. Rev. B} \textbf{67} 174501.
\bibitem{Brydon1} Brydon P M R, Chen W, Asano Y and Manske D 2013 {\it Phys. Rev. B} \textbf{88} 054509.
\bibitem{Brydon2} Brydon P M R, Iniotakis C and Manske D 2009 {\it New J. Phys.} \textbf{11} 055055.
\bibitem{Cheng1} Cheng Q, Jin B and Yu D Y 2015 {\it Phys. Lett. A} \textbf{379} 1172.
\bibitem{Cheng2} Cheng Q and Jin B 2016 {\it Europhys. Lett.} \textbf{113} 17007.
\bibitem{Cheng3} Cheng Q, Zhang K H, Yu D Y, Chen C J, Zhang Y H and Jin B 2016 {\it Supercond. Sci. Technol.} \textbf{29} 075010.
\bibitem{Krive} Krive I V, Gorelik L Y, Shekhter R I and Jonson M 2004 {\it Low Temp. Phys.} \textbf{30} 398.
\bibitem{Dolcini} Dolcini F and Giazotto F 2007 {\it Phys. Rev. B} \textbf{75} 140511(R).
\bibitem{Reynoso} Reynoso A A, Usaj G, Balseiro C A, Feinberg D and Avignon M 2008 {\it Phys. Rev. Lett.} \textbf{101} 107001.
\bibitem{Zazunov} Zazunov A, Egger R, Jonckheere T and Martin T 2009 {\it Phys. Rev. Lett.} \textbf{103} 147004.
\bibitem{Yokoyama} Yokoyama T, Eto M and Nazarov Y V 2014 {\it Phys. Rev. B} \textbf{89} 195407.
\bibitem{Dolcini2} Dolcini F, Houzet M and Meyer J S 2015 {\it Phys. Rev. B} \textbf{92} 035428.
\bibitem{Szombati} Szombati D B, Nadj-Perge S, Car D, Plissard S R, Bakkers E P A M and Kouwenhoven L P 2016 {\it Nature Phys.} \textbf{6} 965.
\bibitem{Qi2} Qi X L, Hughes T L, Raghu S and Zhang S C 2009 {\it Phys. Rev. Lett.} \textbf{102} 187001.
\bibitem{BTK} Blonder G E, Tinkham M and Klapwijk T M 1982 {\it Phys. Rev. B} \textbf{25} 4515.
\bibitem{Eilenberger} Eilenberger G 1968 {\it Z. Phys.} \textbf{214} 195.
\bibitem{Eschrig} Eschrig M 2009 {\it Phys. Rev. B} \textbf{80} 134511.
\bibitem{McMillan} McMillan W L 1968 {\it Phys. Rev.} \textbf{175} 559.
\bibitem{Tanaka4} Tanaka Y and Kashiwaya S 1998 {\it Phys. Rev. B} \textbf{56} 892.
\bibitem{Fu2} Fu L and Kane C L 2009 {\it Phys. Rev. B} \textbf{79} 161408(R).
\bibitem{Crepin} Crepin F, Trauzettel B and Dolcini F 2014 {\it Phys. Rev. B} \textbf{89} 205115.
\bibitem{Kwon} Kwon H J, Sengupta K and Yakovenko V M 2004 {\it Eur. Phys. J. B} \textbf{37} 349.
\bibitem{Beenakker3} Beenakker C W J, Pikulin D I, Hyart T, Schomerus H and Dahlhaus J P 2013 {\it Phys. Rev. Lett.} \textbf{110} 017003.
\bibitem{Furusaki2} Furusaki A and Tsukada M 1991 {\it Solid State Commun.} \textbf{78} 299.


\end{thebibliography}
\end{document}